\documentclass[twocolumn,showpacs,amsmath,amssymb,nofootinbib]{revtex4-1}
\usepackage{graphicx} 
\usepackage{dcolumn}

\begin{document}

\title{First proton-pair breaking in semi-magic nuclei beyond $^{132}$Sn and 
$^{208}$Pb: \\Configuration of the long-lived isomer of $^{217}$Pa}

\author{Alain~Astier}
\author{Marie-Genevi\`eve~Porquet}

\affiliation{
CSNSM, IN2P3-CNRS and Universit\'e Paris-Sud, 
F-91405 Orsay, France}

\date{\hfill \today}

\begin{abstract}
The close similarity between the shell structures in the  
$^{132}$Sn and $^{208}$Pb regions is a well known phenomenon. 
Thus, using the correspondence between the high-$j$ orbits located above the 
$Z=50$ and $Z=82$ shell gaps, we discuss the evolutions of the fully-aligned 
states with one broken proton pair in the $N=82$ and $N=126$ isotones. 
A long-lived isomeric state was discovered in $^{217}$Pa more than thirty years ago
and despite two other experiments giving new experimental results, the discussions on 
its main properties (spin, parity, configuration) remained inconclusive. 
Then, using the comparison with the  $I^\pi=17/2^+$ isomeric state recently measured 
in $^{139}$La, the isomeric state of $^{217}$Pa is assigned as the fully-aligned state
of the $(\pi h_{9/2})^2(\pi f_{7/2})^1$ configuration.
\end{abstract} 

\pacs{21.60.Cs, 23.20.Lv, 27.60.+j, 27.80.+w, } 
 
\maketitle
Isomeric states occupy a prime position in nuclear structure
study. As a general rule, their configuration is nearly
pure, so their discovery enables the identification of single-particle states
among a great number of other excited states of the nucleus. Then their
properties can be compared to theoretical predictions. 
For instance, in odd-$A$ nuclei close to magic numbers, the first long-lived
isomers that had been discovered are due to the M4 character 
of the isomeric transitions~\cite{fe49,no49}, which is explained by the difference in 
angular momentum of two orbits close in energy~\cite{ma50}, such as
$p_{1/2}-g_{9/2}$ for $N(Z) < 50$, $d_{3/2}-h_{11/2}$ for $N(Z) < 82$, 
and $f_{5/2}-i_{13/2}$ for $N < 126$. 
Another mechanism is known to produce isomeric states in spherical nuclei close to 
magic numbers, the breaking of a high-$j$ nucleon pair. When the angular
momenta of the two nucleons are fully aligned, the overlap
of their wavefunctions is maximized and their residual interaction is large.
Thus the excitation energy of the 
state having the maximum spin ($I_{max}=2j-1$) is lowered and 
its $E2$ decay towards the next state of the multiplet is slowed
down. Textbook
examples are the 8$^+$ isomeric state ($T_{1/2}$= 201~ns) in $^{210}$Pb 
and the 8$^+$ isomeric state ($T_{1/2}$= 99~ns) in $^{210}$Po, 
where the two nucleons outside the douby-magic core are located in 
the $\nu g_{9/2}$ and $\pi h_{9/2}$ orbits, respectively. 
Such isomeric states due to breaking of a high-$j$ nucleon pair are also 
found in semi-magic nuclei having much more than two valence nucleons, for
instance in the $N=82$ isotones with $54 < Z < 60$ and in the $N=126$ isotones
with $85 < Z < 92$~\cite{nndc}. 

In the nuclear chart, the comparison of the regions lying above the two 
doubly-magic nuclei, $^{132}$Sn 
and $^{208}$Pb, is very interesting since each single-particle
state in the $^{132}$Sn region can be related to one particular 
state in the $^{208}$Pb region which only
differs by one unit of orbital angular momentum~\cite{bl81}. These 
corresponding states have approximately the same ordering and spacing, 
so one may expect similarities in nuclei belonging to these two
regions.
In this paper, we discuss the properties of the fully-aligned states with 
one broken proton pair involving the high-$j$ orbits located above the $Z=50$ 
gap in the $N=82$ isotones with $52 \le Z \le  58$. The isomeric 
state recently identified in
$^{139}_{57}$La is particularly emphasized. Then the comparison with the  
fully-aligned states measured in the $N=126$ isotones having 
$84 \le Z \le  92$ allows us to assign the
configuration of the long-lived isomeric state of $^{217}$Pa, for which no
firm conclusion was provided in the previous papers. 

The high-spin structures of five $N=82$ isotones ($54 \le Z \le 58$) have 
been recently studied~\cite{as12}. Positive-parity states dominate the 
low-energy part of the level schemes, since the two proton orbits located just 
above the $Z=50$ gap are $\pi g_{7/2}$ and $\pi d_{5/2}$. In $^{133}_{51}$Sb,
the latter is located 962~keV above the former. Even though the pairing
correlations dilute the occupancy of the two subshells when $Z$ is increasing,
the crossing of the two orbits occurs at the 
'right' place, the ground state of $^{139}_{57}$La having $I^\pi=7/2^+$ and the 
one of $^{141}_{59}$Pm having $I^\pi=5/2^+$. 

\begin{figure*}[!t]
\begin{center}
\includegraphics*[width=14cm]{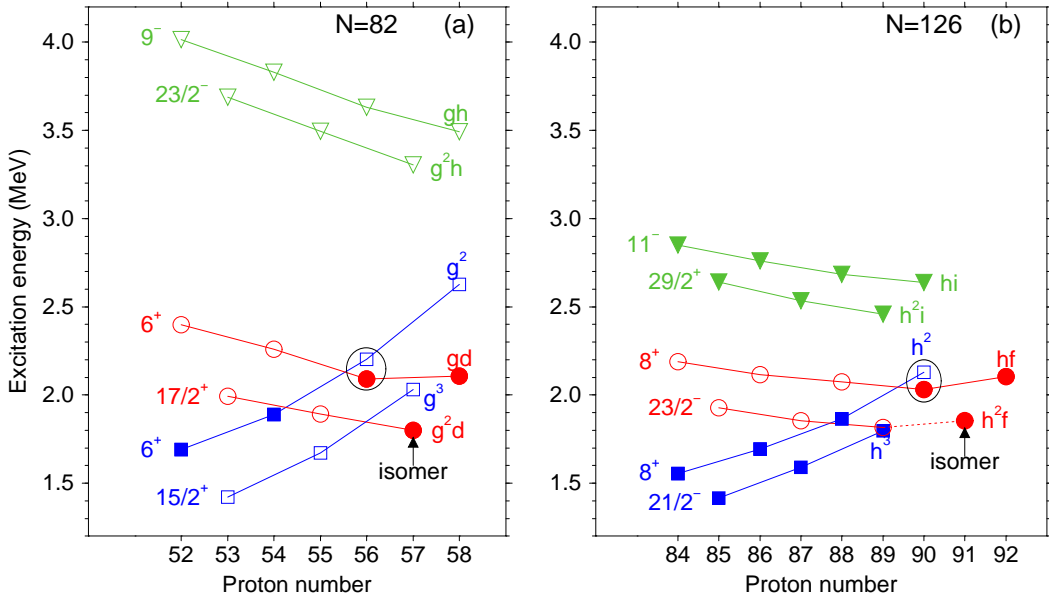}
\caption{(Color online) Experimental excitation energies of the fully-aligned states with one broken
proton pair in the $N=82$ isotones with $52 \le Z \le 58$, involving the three
high-$j$ orbits located just above the $Z=50$ gap, $\pi g_{7/2}$, $\pi d_{5/2}$,  and 
$\pi h_{11/2}$ (a) and in the $N=126$ isotones with $84 \le Z \le 92$, involving the three
high-$j$ orbits located just above the $Z=82$ gap, $\pi h_{9/2}$, $\pi f_{7/2}$,  and 
$\pi i_{13/2}$ (b). The isomeric states are drawn with filled symbols.
The ellipse drawn at $Z=56(90)$ means that the two 6$^+$(8$^+$) 
states are strongly mixed. The low-lying isomeric states measured in 
$^{137}_{57}$La and $^{217}_{91}$Pa and discussed in this paper are marked.}
\label{seniorite2et3}       
\end{center}
\end{figure*} 

All the even-$Z$ isotones having $52 \le Z \le 58$ exhibit a low-lying isomeric 
state with $I^\pi=6^+$. It is due to the {\it first} breaking of a proton pair which
has either the $(\pi g_{7/2})^2$ configuration  or the 
$(\pi g_{7/2})^1(\pi d_{5/2})^1$ one, since both configurations lead to $I^\pi_{max}=6^+$
and display a small gap in energy between the 6$^+$ and the 4$^+_1$ states.
Indeed two 6$^+$ states have been identified in these isotones~\cite{zh96,as12,nndc}, their configurations 
depend  on the location of the Fermi level within the two orbits. For $Z=52$ and 54, 
the 6$^+_1$ state has most likely the first configuration and the 6$^+_2$ state the
second one, given that the $\pi g_{7/2}$ orbit is filling first. 
On the other hand, the results of 
($^3$He, $d$) reactions~\cite{nndc} clearly indicate that (i) the 6$^+_1$ state of
$^{140}_{58}$Ce comes from the $(\pi g_{7/2})^1(\pi d_{5/2})^1$ configuration and its  
6$^+_2$ state from the  $(\pi g_{7/2})^2$ one, and (ii) the two 6$^+$ states of
$^{138}_{56}$Ba are strongly mixed. It is worth noting that, while the  6$^+_1$ state 
of these four isotones is isomeric, whatever its configuration, the decay of their 
6$^+_2$ state is never isomeric, as the $M1$ transition to
the 6$^+_1$ state is never delayed.

As to the odd-$Z$ isotones, the result of the {\it first} breaking of a 
proton pair varies with the occupation of the two
orbits. If the odd proton is located in the $\pi g_{7/2}$ orbit, this leads 
to the $(\pi g_{7/2})^{\pm3}$ configuration or to the 
$(\pi g_{7/2})^{\pm1}(\pi d_{5/2})^2$ one, depending on the total number of protons. 
It is important to note that both configurations give $I^\pi_{max}=15/2^+$. 
When the odd proton is 
promoted to the $\pi d_{5/2}$ orbit, the configuration is 
$(\pi g_{7/2})^2(\pi d_{5/2})^1$, with $I^\pi_{max}=17/2^+$. As expected, the 
$(\pi g_{7/2})^{\pm3}$ configuration gives rise to the the first yrast 
states of $^{135}_{53}$I and $^{137}_{55}$Cs, with $I^\pi=11/2^+$ and 15/2$^+$. The 
$(\pi g_{7/2})^2(\pi d_{5/2})^1$ configuration lies at higher excitation energy and 
only the state with $I^\pi_{max}=17/2^+$ belongs to their yrast sequence~\cite{zh96,as12}.
Unlike the even-$Z$ isotones, neither $^{135}$I nor $^{137}$Cs exhibit a low-lying 
isomeric state, as the distance in energy between the 15/2$^+$  and 11/2$^+$ 
states of the $(\pi g_{7/2})^{3}$ configuration is well higher than the one between 
the 6$^+$ and 4$^+$ states of the $(\pi g_{7/2})^2$ configuration 
(compare Fig.~12 and Fig.~13(a) of Ref.~\cite{as12}).  

$^{139}_{57}$La behaves differently from $^{135}_{53}$I and $^{137}_{55}$Cs~\cite{as12}: (i) two coexisting 
structures have been identified, one built on the 7/2$^+$ ground state and another 
one built on the first-excited state at 166~keV having $I^\pi=5/2^+$; (ii) 
the 17/2$^+_1$ state is isomeric, with $T_{1/2}$~=~315~ns. The first structure  
contains four states with spins 7/2$^+$, 11/2$^+$, 13/2$^+$ and 15/2$^+$. The second 
structure also contains four states, with spins 5/2$^+$,  9/2$^+$, 13/2$^+$ and 17/2$^+$. 
Results of the shell model calculations done in Ref.~\cite{as12} indicate that the
states of the two groups have not the same configuration. The main configuration of 
all the
states of the first structure is $(\pi g_{7/2})^5(\pi d_{5/2})^2$, i.e.,
the odd proton occupies the $\pi g_{7/2}$ orbit. On the other hand, the main 
configuration of all the states of the second structure is 
$(\pi g_{7/2})^6(\pi d_{5/2})^1$, i.e., the odd proton occupies the $\pi d_{5/2}$ orbit.
The relative energies of the states of these two structures explain why the 
17/2$^+_1$ state is isomeric in $^{139}$La. Unlike $^{135}$I and $^{137}$Cs, it 
cannot be linked to the 15/2$^+$ state which is located above it, therefore it 
decays towards the 
13/2$^+_1$ state by a low-energy $E2$ transition.

\begin{figure*}[!htb]
\begin{center}
\includegraphics[angle=90,width=18cm]{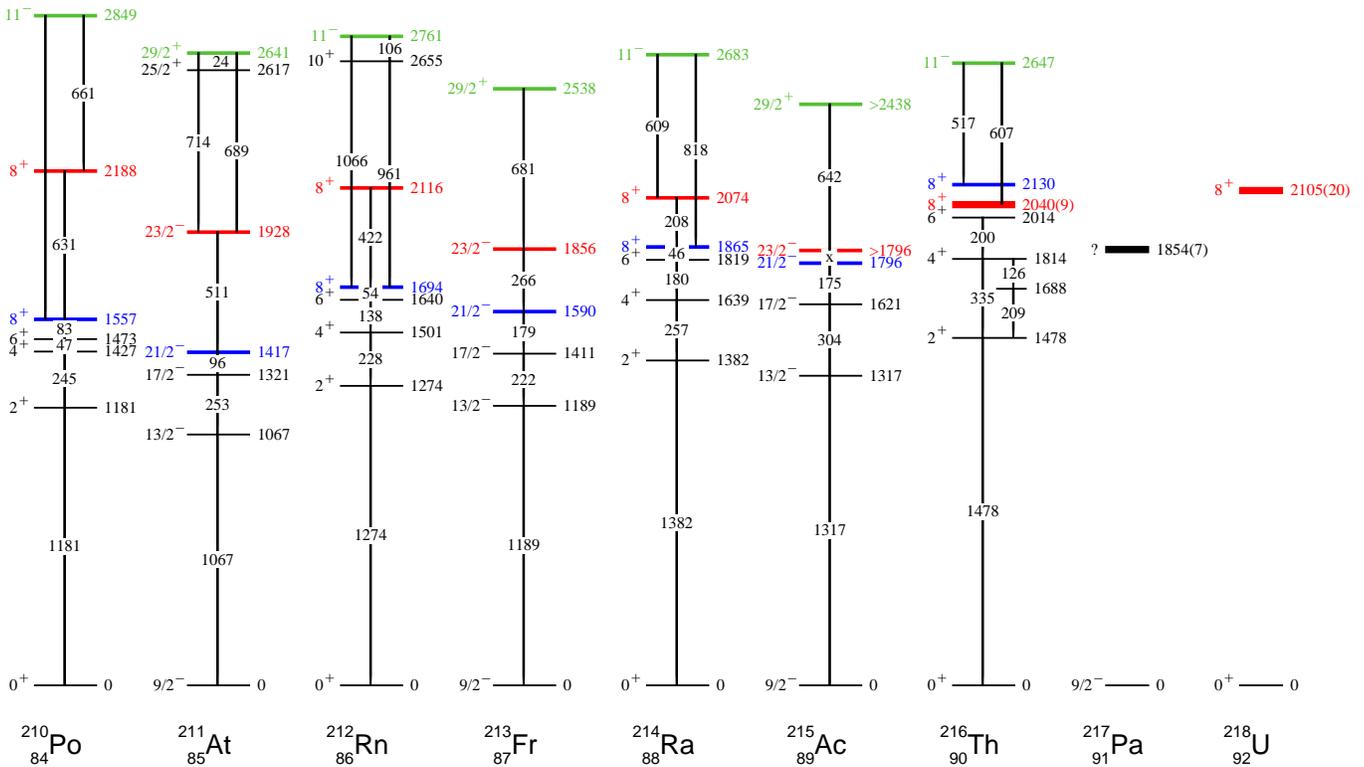}
\caption{(Color online)  
Systematics of the first excited states of the $N=126$ isotones with $84 \le Z \le 92$, 
the colored states are the fully-aligned states with one broken proton pair shown 
in Fig.~\ref{seniorite2et3}(b). The $g$ factors of most of the 8$^+_1$, 11$^-$, 
21/2$^-$, and 29/2$^+$ states have been measured, confirming their proton
configurations (see for instance, the figures 4(a) and 4(b) of
Ref.~\cite{de83}).
The $\alpha$-decaying states of $^{216}$Th, $^{217}$Pa 
and $^{218}$U are drawn with a thicker line.
The experimental data come from Refs.~\cite{de83,he02,le07} for
$Z=89$, 91 and 92, respectively and from the 'Evaluated Nuclear Structure Data File'~\cite{nndc} 
for the others. 
}
\label{levelschemes}       
\end{center}
\end{figure*}

Fig.~\ref{seniorite2et3} illustrates the close similarity between the shell structures 
in the $^{132}$Sn and $^{208}$Pb regions. As mentioned above, 
the orbital angular momenta differ by one unit, the orientation of the 
intrinsic spin remaining unchanged, namely, 
$\pi g_{7/2} \rightarrow  \pi h_{9/2}$, $\pi d_{5/2} \rightarrow \pi f_{7/2}$,  and 
$\pi h_{11/2} \rightarrow \pi i_{13/2}$. 
In Fig.~\ref{seniorite2et3}(a), we have gathered the 
results discussed above, i.e., the evolution of the excitation energy of the 
fully-aligned states with one broken proton pair in the $N=82$ isotones with 
$52 \le Z \le 58$. For the sake of completeness, the states 
involving the third high-$j$ orbit located above the $Z=50$ gap, $\pi h_{11/2}$, 
are also shown (the experimental data are from Refs.~\cite{zh96,as12}). \\
Fig.~\ref{seniorite2et3}(b) shows the analogous excited states measured in the $N=126$ isotones with 
$84 \le Z \le 92$ (the low-energy parts of their level schemes are shown in
Fig.~\ref{levelschemes}). 
The two 8$^+$ states of
the  $N=126$ isotones display the same evolution as the two 6$^+$ states of
the  $N=82$ isotones. Similarly the behaviors of the 21/2$^-$ and 23/2$^-$ states of the 
$N=126$ isotones are the same as the ones of the 15/2$^+$ and 17/2$^+$ states of the 
$N=82$ isotones. Thus the long-lived isomeric state of $^{217}_{91}$Pa, which was 
located at $\sim$1854(7)~keV by means of its $\alpha$ decay, is very likely the 
counterpart of the isomeric state of $^{139}$La.

Before going futher, it is important to recall 
the experimental status of this isomeric state.
The alpha decay of an isomeric state of $^{217}$Pa was 
observed many years ago from the $^{40}$Ar+$^{181}$Ta reaction 
at 176~MeV bombarding energy~\cite{sc79}. Compared with the
information known at that time in the lighter $N=126$ isotones, this
isomer was associated with a fully-aligned three-proton state, 
either with $I^\pi=29/2^+$ from the $(\pi h_{9/2})^2(\pi i_{13/2})^1$ 
configuration or with $I^\pi=23/2^-$ from the
$(\pi h_{9/2})^2(\pi f_{7/2})^1$ one. Its very long half-life ($T_{1/2}=$1.6~ms)
was proposed to be due to a large hindrance from the angular momentum
of the emitted alpha particle. Many years later, this
isomeric state was produced in another reaction,  
$^{28}$Si+$^{194}$Pt at a beam energy of 163~MeV~\cite{ik98}, and the 
large hindrance of the transition was entirely attributed to the
centrifugal barrier, a value of $\ell \sim 9\hbar$ being
estimated from standard formula. Given that the ground state 
of the daughter nucleus, $^{213}$Ac, has $I^\pi=9/2^-$, the isomeric
state of $^{217}$Pa was thus assigned as the fully-aligned state of 
the $(\pi h_{9/2})^2(\pi i_{13/2})^1$ configuration, with 
$I^\pi=29/2^+$. Later on, a more detailed study of the decay scheme of
$^{217}$Pa was performed by means of $\alpha-\gamma$
spectroscopy~\cite{he02}. Fine structures in the $\alpha$ decay of the
isomeric state as well as of the ground state of $^{217}$Pa were 
observed, leading to the unambiguous identification of several excited states
of $^{213}$Ac. Moreover, the measured decay curve of the main 
$\alpha$ transition decaying the ground state of $^{217}$Pa was found 
to be composite, indicating that the isomeric state 
($T_{1/2}=$1.2(2)~ms) partly decays to
the ground state, \%IT=27(4), giving the partial half-life,
$T_{1/2}^\gamma \sim 4.4$~ms. This last result was used to provide
support to the $I^\pi=29/2^+$ choice~\cite{he02}: The isomeric transition
was attributed to a $29/2^+ \rightarrow 23/2^-$ decay, i.e., an 
$E3$ transition with an energy low enough to be in agreement with a     
long half-life. 

Nevertheless, such an interpretation is clearly at variance with the behavior of the 
neighboring isotones, where the excitation energies of the fully-aligned states 
involving the $\pi i_{13/2}$ subshell decrease slowly as a function of $Z$ 
[see the evolution of the 11$^-$ and 29/2$^+$ states drawn in green  
in Fig.~\ref{seniorite2et3}(b) and Fig.~\ref{levelschemes}]. A drop of about 650~keV of the 
29/2$^+$ state from $^{215}_{89}$Ac to $^{217}_{91}$Pa is very unlikely, 
especially since the evolution of the 11$^-$ state of the even-$Z$ 
remains smooth up to $Z=90$.
On the other hand, the excitation energy of the long-lived isomeric state of 
$^{217}$Pa fits very well the evolution of the fully-aligned 23/2$^-$ state, 
which is expected at lower energy than the 21/2$^-$ state from the $(\pi h_{9/2})^3$
configuration at $Z=91$, given their opposite evolution as a function of $Z$ 
[see Fig.~\ref{seniorite2et3}(b)]. 
  
Thus the isomeric states of $^{139}$La and $^{217}$Pa share 
several similar properties, 
(i) their excitation energies are close to each other, (ii) their 
configurations involve the two first orbitals located just above the proton
shell gaps, differing by one unit of orbital angular momentum, namely 
$(\pi g_{7/2})^2(\pi d_{5/2})^1$ ($I^\pi_{max}=17/2^+$) for the former and 
$(\pi h_{9/2})^2(\pi f_{7/2})^1$ ($I^\pi_{max}=23/2^-$) for the latter.

On the other hand, the half-live of their gamma decays are very different. 
The half-life of the isomeric 
state of $^{139}$La is 315~ns, which is due to the emission of an $E2$ 
transition of 89~keV with $B(E2)=$1.9~W.u.~\cite{as12}, value in line with
the fact that both the emitting and populated states have the same
configuration, $(\pi g_{7/2})^6(\pi d_{5/2})^1$. 
The partial half-life of the isomeric state of $^{217}$Pa is 
$T_{1/2}^\gamma \sim 4.4$~ms.
Such a long half-life cannot be due to  
a low-energy $E2$ transition between two states having the same configuration, 
implying that the 19/2$^-$ state
belonging to the $(\pi h_{9/2})^2(\pi f_{7/2})^1$ multiplet has to be located
\emph{above} the 23/2$^-$ state.
This is in agreement with results of 
large-scale shell model calculations which were performed for the $N=126$ 
isotones, from $Z=84$ to $Z=94$~\cite{ca03}. 

As said above, an $E3$ transition with a low energy can lead to a partial 
half-life in the ms range. For instance, for $E_\gamma = $20~keV and 
$T_{1/2}^\gamma \sim 4.4$~ms, the value of the reduced transition
probability is $B(E3)\sim 30~W.u.$. Then the $E3$ isomeric transition of 
$^{217}$Pa would link the 23/2$^-$ fully-aligned 
state of the $(\pi h_{9/2})^2(\pi f_{7/2})^1$ configuration to a 
17/2$^+$ state. Given that in $^{217}$Pa, the $\pi i_{13/2}$ subshell is expected 
at an energy about 600~keV by extrapolating the values known in the lighter 
isotones~\cite{nndc}, the 17/2$^+$ state belonging to 
the $(\pi i_{13/2})^1(\pi h_{9/2})^2$ multiplet could be located below 
$\sim$~1850~keV, the energy of the long-lived isomeric state.  
Moreover it is worth recalling that the same configurations are involved in 
the $E3$ transitions measured in the 
lighter isotones ($^{211}$At, $^{213}$Fr and $^{215}$Ac~\cite{nndc}) and their  
$29/2^+ \rightarrow 23/2^-$ transitions have $B(E3)\sim 25~W.u.$, close to
the value mentioned above for $^{217}$Pa. 

As to the large hindrance of the $\alpha$ decay, 
$^{217}$Pa$^m$~$\Rightarrow$~$^{213}$Ac$^{gs}$, it is likely
due to (i) the centrifugal barrier (the angular momentum of the emitted $\alpha$
has to be at least 8~$\hbar$ for a $23/2^- \rightarrow 9/2^-$ transition) and 
(ii) the change in proton configuration (the proton pair of the $\alpha$
particle has to be formed from two protons lying in two different orbits and
having  their angular momentum almost aligned,  
$[(\pi h_{9/2})^2(\pi f_{7/2})^1]_{23/2^-} \rightarrow 
[(\pi h_{9/2})^1]_{9/2^-}$).

In summary, using the correspondence between the high-$j$ orbits located above the $Z=50$
and $Z=82$ shell gaps, as well as the $I^\pi=17/2^+$ isomeric state 
recently measured in $^{139}$La, we propose that the long-lived isomeric 
state of $^{217}$Pa, which was discovered more than thirty years ago, is the
fully-aligned state with $I^\pi=23/2^-$ of the 
$(\pi h_{9/2})^2(\pi f_{7/2})^1$ configuration, its gamma partial half-life 
being in agreement with an $E3$ transition having an energy of about 20~keV. 
It is worth recalling that the $g$ factors of all the high-spin isomeric states of the
$N=126$ isotones [drawn with filled symbols in 
Fig.~\ref{seniorite2et3}(b)] have been measured for $Z \leq 89$, confirming the identification of their
two-proton configurations. Unfortunately the measurement of the $g$ factor of $^{217}$Pa$^m$ seems
out of reach by now, both its very low production rate in reactions induced by
heavy ions and its long lifetime prevent use of
the standard methods available in the present experimental facilities. 

Finally, one may wonder whether such a mechanism giving rise to long-lived isomeric 
states may exist elsewhere in the nuclear chart. Starting with the $^{78}$Ni region, 
we find the $\pi f_{5/2}$ and $\pi p_{3/2}$ orbits lying above the 
$Z=28$ shell gap. Thus the three-proton isomeric state would be the fully-aligned state of the  
$(\pi f_{5/2})^2(\pi p_{3/2})^1$ configuration with $I^\pi_{max} = 11/2^-$, provided that it is located 
below the 9/2$^-$ state of the $(\pi f_{5/2})^3$ configuration. 
We know that it is not the case, since the yrast states of  $^{83}_{33}$As 
have been recently identified and the 
11/2$^-$ state, higher in energy, decays to the 9/2$^-$ state by means of a $M1$
transition of 323~keV~\cite{po11}.

On the other hand, long-lived isomeric states from three-neutron configurations could be 
found in heavy-$A$ Pb isotopes. Indeed high-spin states were observed in
$^{211}$Pb$_{129}$ having three valence neutrons in the $\nu g_{9/2}$ and 
$\nu i_{11/2}$ orbits which are lying just above the $N=126$ shell
closure~\cite{la05}. The fully-aligned states of the $(\nu g_{9/2})^3$ 
and $(\nu g_{9/2})^2(\nu i_{11/2})^1$ configurations (with $I^\pi=21/2^+$ and 
27/2$^+$, respectively) are isomeric, because of the low energy of their decaying $E2$ 
transition. As stated in
the conclusion of Ref.~\cite{la05}, predictions can be made for the expected behavior
of more neutron-rich Pb isotopes: {\it "The  27/2$^+$ level will continue to be isomeric,
either an $E2$-decaying isomer as in $^{211}$Pb or, should it move below the 23/2$^+$
state, very long lived."}  
Thus, even though the order of the two valence orbits is reversed as compared to the   
$N=82$ and $N=126$ isotones discussed above (here, the low-$j$ orbit lies above the  
high-$j$ one), the 
heavy-$A$ Pb isotopes could display an yrast trap with $I^\pi=27/2^+$, from the
$(\nu g_{9/2})^2(\nu i_{11/2})^1$ configuration. 

When moving to the super-heavy part of the chart, another
interesting case may be also found. Different spherical doubly-magic super-heavy nuclei are predicted,
depending on the parametrization of the mean field~\cite{be99}. Nevertheless the
existence of a
spherical gap at $N=184$ seems to be almost 'universal'. The two orbits lying just 
above this gap are $\nu j_{13/2}$ and $\nu h_{11/2}$, i.e., with $\Delta \ell = 2$ and
$\Delta j = 1$, as all the couples of orbits discussed above. Thus an isomeric 
state with $I^\pi = 35/2^-$
from the $(\nu j_{13/2})^2(\nu h_{11/2})^1$ configuration would be expected 
in the $Z=114(120)$ isotopes having $N > 187$, provided that such nuclei 
remain spherical when adding neutrons in the two orbits.  A very large hindrance 
of the $\alpha$ decay of such a 35/2$^-$ state is expected because of the centrifugal barrier
($\ell > 11\hbar$ when considering its decay to the ground state of the daughter 
nucleus). Therefore the half-life of the isomeric state would be well longer than the one of the ground
state (the same situation as the 18$^+$ isomeric state of $^{212}$Po, with $T_{1/2}$~=~45~s, as
compared to $T_{1/2}^{gs}~=~0.299~\mu$s~\cite{nndc}).


\end{document}